\documentclass[reprint,
superscriptaddress,aps,pre,twocolumn,floatfix,showpacs,amsmath,longbibliography,10pt]{revtex4-2}

\usepackage{booktabs, float}
\usepackage{xcolor}
\usepackage{graphicx}
\usepackage{graphics}
\usepackage{amsmath}
\usepackage{amssymb}
\usepackage{amsthm,amstext}
\usepackage{amsfonts}
\usepackage{lipsum}
\usepackage{MnSymbol}
\usepackage{mathrsfs}
\usepackage{stmaryrd}
\usepackage{latexsym}
\usepackage[applemac]{inputenc}
\usepackage{accents}

\usepackage{soul}
\usepackage{xcolor}

\newcommand{\md}[1]{\textcolor{black}{#1}}

\usepackage[bbgreekl]{mathbbol}
\DeclareSymbolFontAlphabet{\mathbbm}{bbold}
\DeclareSymbolFontAlphabet{\mathbb}{AMSb}
\DeclareMathAlphabet\mathbfcal{OMS}{cmsy}{b}{n}

\renewcommand\d\delta
\newcommand\D\Delta

\usepackage[normalem]{ulem}

\begin{document}

\title{Canceling  the elastic Poynting effect with geometry}

\author{M. Destrade$^{\dagger}$, Y. Du$^{\dagger\dagger *}$, J. Blackwell$^{\dagger}$, N. Colgan$^{\dagger}$, V. Balbi$^{\dagger}$}
\affiliation{\it$^{\dagger}$School of Mathematical and Statistical Sciences, University of Galway, Galway, Ireland\\ 
$^{\dagger\dagger}$Department of Mathematics and Statistics, University of Glasgow, Glasgow, Scotland.}

\date{\today}


\begin{abstract}

The Poynting effect is a paragon of nonlinear soft matter mechanics. 
It is the tendency (found in all incompressible, isotropic, hyperelastic solids) exhibited by a soft block to expand vertically when sheared horizontally. 
It can be observed whenever the length of the cuboid is at least four times its thickness. 
Here we show that the Poynting effect can be easily reversed and the cuboid can shrink vertically, simply by reducing this aspect ratio.
In principle, this discovery means that for a given solid, say one used as a seismic wave absorber under a building, an optimal ratio exists where vertical displacements and vibrations can be completely eliminated.
Here we first recall the classical theoretical treatment of the positive Poynting effect, and then show experimentally how  it can be reversed. Using Finite Element simulations, we then investigate how the effect can be suppressed.
\md{We find that cubes always provide a reverse Poynting effect, irrespective of their material properties (in the third-order theory of weakly nonlinear elasticity).}

\end{abstract}


\maketitle


\section{Introduction} 

The importance and ubiquity of the simple shear deformation is well recognised in the natural and man-made worlds. 
For example, it is at play both in earthquakes, when tectonic plates move in opposite directions along a fault, and in earthquake protection, when seismic base isolation systems decouple buildings from ground motion. 
Simple shear is also an accepted standard protocol for testing the mechanical properties of elastic solids. 
The Poynting effect in simple shear is an aspect which might be less appreciated because it is a nonlinear effect which occurs beyond the realm of linear elasticity. 

In 1909, Poynting \cite{poynting1909pressure} observed that long cylindrical steel rods lengthen when twisted. 
Four decades later, Rivlin \cite{rivlin1947torsion} proved this result analytically in a landmark paper of nonlinear rubber elasticity. 
He went on to show that rubber tends to expand vertically when sheared horizontally \cite{rivlin1948large}.
Because of this effect, buildings resting on base isolations systems can experience vertical forces in principle. 
Also, when the skull is subjected to a rotational acceleration as in a boxing event, brain matter may experience not only shearing forces in the horizontal plane but also vertical forces of a magnitude compatible with axonal diffusion injury \cite{balbi2019poynting}. 
Here we show that the effect may be reduced, and even reversed, with choices of the solid's geometry.

The Poynting effect in simple shear is easy to prove in the framework of exact, incompressible, isotropic elasticity, where the strain energy density $W$ depends on two strain invariants only, $I_1=\text{tr}(\mathbf F^T\mathbf F)$, $I_2=\text{tr}(\mathbf F^{-1}\mathbf F^{-T})$, and $\mathbf F$ is the deformation gradient. 
Indeed Rivlin found that when a solid is sheared by an amount $K$, simple shear can be modelled by the deformation 
\begin{equation}
x_1=X_1+KX_2, \quad x_2=X_2, \quad x_3=X_3,
\label{simple-shear}
\end{equation}
bringing a particle originally at $\mathbf X$ to its current position at $\mathbf x$.
Assuming there is no stress applied in the out-of-plane direction, the deformation is maintained by the application of the following Cauchy stress
\begin{equation}
\sigma_{11} = 2\frac{\partial W}{\partial I_1}K^2, \;  
\sigma_{22} = -2\frac{\partial W}{\partial I_2}K^2, \;  
\sigma_{12} = 2\left(\frac{\partial W}{\partial I_1}+\frac{\partial W}{\partial I_2}\right)K, 
\end{equation}
while all the other components are zero. 
Clearly, $\partial W/\partial I_2 \ne 0$ in general, showing that it is necessary to provide a normal force to effect the simple shear. 

Now consider the \textit{Mooney-Rivlin model} $W=C_1(I_1-3)/2 + C_2(I_2-3)/2$, where $C_1$, $C_2$ are positive constants.
Then $\sigma_{22} = -C_2K^2$, showing first, that the Poynting effect is not captured by linear elasticity (where quadratic terms are neglected) and second, that the solid will expand vertically unless a force is applied downwards to compensate $\sigma_{22}$.
Note that this result is actually valid for all solids for small-to-moderate amounts of shear because, as noted by Rivlin, the Mooney-Rivlin model coincides \cite{rivlin1951large, destrade2010onset}, at the same level of approximation, with the most general model of (incompressible, isotropic) third-order elasticity.

The idealised deformation \eqref{simple-shear} is for a solid of infinite extend. 
In practice, we must deal with finite dimensions. 
Typically, in the lab we prepare a cuboid of length $L$, thickness $H$ and width $A$, and glue two rigid platens to its two large faces. 
We then move one platen parallel to the other by a distance $d$ (see Fig.~\ref{figure1}), so that $K=d/H$, and measure the horizontal component $F$ of the required force to  obtain the shear stress component experimentally as $\sigma_{12} = F/(AL)$.

A major conceptual issue arises then, as it turns out that traction forces must also be applied to the slanted faces to prevent them from bending and realise the homogeneous deformation \eqref{simple-shear}.
These forces are \textit{never} applied in the lab. 
Instead, the accepted standard protocol \cite{ISO1827} is to use ``thin'' cuboids such that $H/L<0.25$, hence minimising the effects of face bending to be of local and small magnitude. 

%
%
\section{Results} 

\begin{figure}[H]
\includegraphics[width=0.45\textwidth]{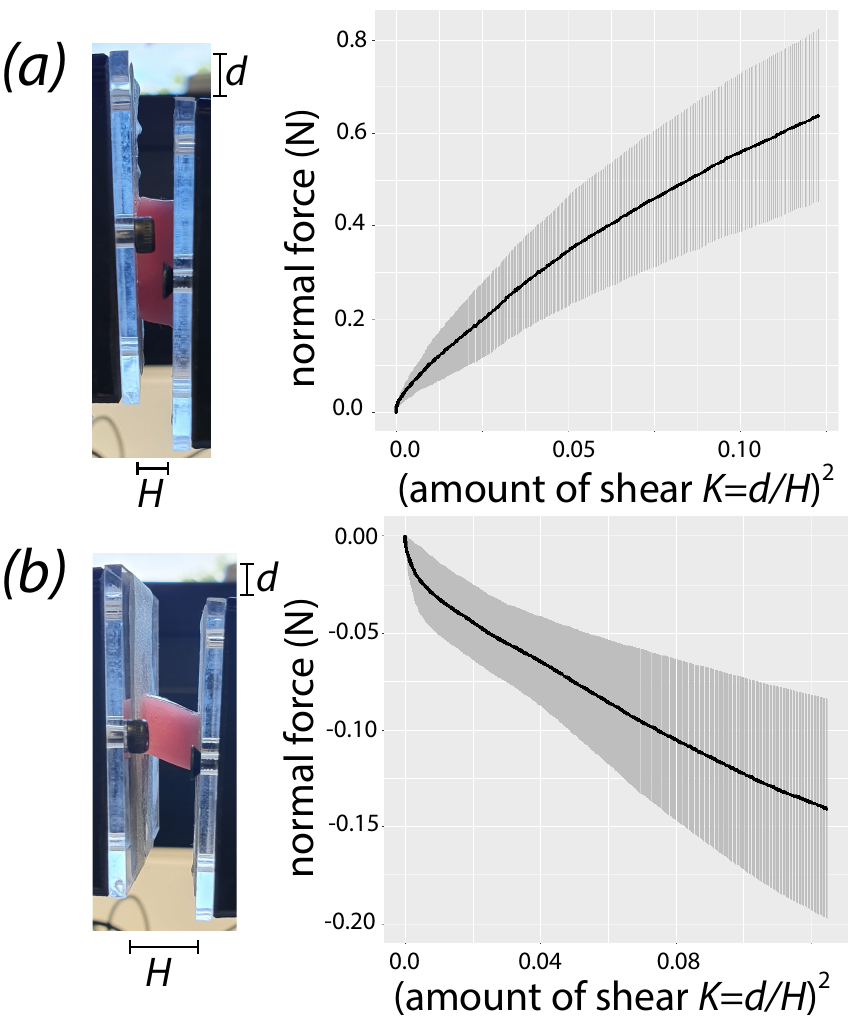}
\caption{Experimental shear of a soft block. (a) When the length/height ratio is large, the cuboid tends to expand in thickness and exerts a positive normal force on the shearing platens (positive Poynting effect). (b) When it is close to unity, the Poynting effect is reversed.}
\label{figure1}
\end{figure}

Fig.~\ref{figure1} displays our main result, showing that the Poynting effect can be reversed by decreasing the aspect ratio of the block.

Fig.~\ref{figure1}(a) shows the shear of a rectangular cuboid ($20 \times 20 \times 5$ mm, $H/L=0.25$) made of silicone.
The data ($n=9$ samples, with standard deviation on each side of the average) reveals an almost linear relationship between the normal force and the squared amount of shear for $K$ up to $0.35$ ($K^2$ up to $0.125$), confirming that the third-order/Mooney-Rivlin modelling is adequate. 
Here the normal force is positive, and the cuboid pushes against the platens. 
In contrast,   Fig.~\ref{figure1}(b) reveals that when cubic samples ($10 \times 10 \times 10$ mm, $H/L=1.0$, $n=8$) are sheared, the normal force is negative, because the geometry leads to large variations in the distribution of stresses and strains: the homogeneous simple shear \eqref{simple-shear}  is no longer a valid approximation of  the deformation taking place, leading to a reversal of the Poynting effect.
For consistency across both pictures, we plot the normal force against $K^2$, and we use $F$ instead of $\sigma_{12}$ because the stress cannot be given in (b), as it varies across the surface of the platens. 
We now anticipate that between the ratios $H/L=0.25$ and $H/L=1.0$ there might exist a geometry for which the normal force is close to zero.

We write the Mooney-Rivlin constants as $C_1=\mu(1-\beta)/2$, $C_2=\mu(1+\beta)/2$, where $\mu$ is the infinitesimal shear modulus and $\beta$ takes any value in $[-1,1]$ to cover the gamut of all Mooney-Rivlin solids or, equivalently for small-but-finite deformations, all third-order elastic solids. 
The theory based on the homogeneous deformation \eqref{simple-shear} predicts a normal force of magnitude $-(\mu/2)(1 + \beta)K^2$ per unit area. 
Here, using Finite Element Analysis, we compute the normal force $N$ to be applied on the top face to maintain the horizontal displacement of the platen. 
\md{We note that alternatively, an analytical, asymptotic treatment is possible \cite{wang2014generalized}; however, it requires the introduction of torques for the sheared faces, which we were not able to measure in our experiments.}

\begin{figure}[H]
\includegraphics[width=0.45\textwidth]{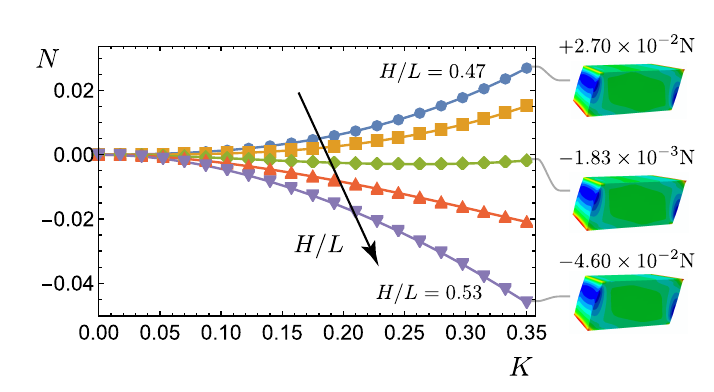}
\caption{Computed normal force $N$ in the simple shear of a soft block with $\beta=0.3$, \md{cross-section $20 \times 20$ mm}, and aspect ratios $H/L = \md{0.47, 0.48, 0.495, 0.51, 0.53}$ \md{(see arrow)}. When \md{$H/L = 0.495$, $N$ is one order of magnitude smaller than when $H/L=0.48, 0.51$ and we conclude that the Poynting effect is effectively cancelled when $H/L \simeq 0.495$}.}
\label{figure2}
\end{figure}

\begin{figure}[H]
\includegraphics[width=0.45\textwidth]{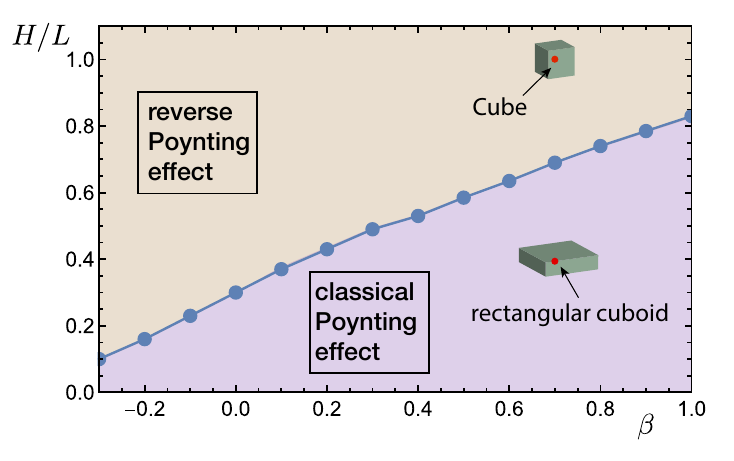}
\caption{Plot of the critical aspect ratio $H/L$ of cancelled Poynting effect.
\md{The two points (see arrows) correspond to the characteristics of the cubic and rectangular cuboid samples used in the experiments.}}
\label{figure3}
\end{figure}

Figure \ref{figure2} illustrates our strategy when $\beta = 0.3$ and $H/L$ varies between \md{0.47 and 0.53}. 
We see that the computed normal force switches from a \md{positive} value (classical Poynting effect) to a \md{negative} value (reverse Poynting effect) as $H/L$ increases, and is almost zero (canceled Poynting effect) when $H/L = \md{0.495}$, giving one point in our $H/L$ Vs $\beta$ graph displayed on Figure \ref{figure3}. 

Figure \ref{figure3} shows the resulting graph, of the critical $H/L$ ratio for all materials ($-0.3 \le \beta \le 1.0$). 
Practically, a block of a given material parameter $\beta$ and given geometry $H/L$, for which the ($\beta, H/L$) point is below (respectively, above) that curve, will expand (respectively, shrink) in simple shear.
The curve itself provides the required block geometry for the annihilation of the normal force. 
\md{We note that cubes ($H/L=1.0$) always produce the reverse Poynting effect, irrespective of their material parameters.}

\md{For further illustration, we used Finite Element simulations (not shown) to determine the material parameters of our samples through inverse analysis.
We found $\mu=62.5$ kPa and $\beta \simeq 0.7$, which allowed us to place the corresponding experimental points on the figure.}

\section{Materials and methods}

We created an experimental setup similar to that developed by Yan et al.~\cite{yan2021novel}. 
A 100 N (accuracy $<0.5\%$) load cell was attached to a Zwick Roell Tensile Tester (Zwick Roell Ltd, UK) with an additional 500 N (accuracy $<0.5\%$) load cell (Richmond Industries, UK) attached to a laptop PC. 
It was not possible to record both data concurrently on the same PC; instead, the start times were synced to within 0.1 s using internal timers. Both load cells were calibrated within two weeks of testing. 

A rig was designed so that samples were sheared as the tensile tester moved vertically. 
The 100 N load cell measured the shear force and the 500 N load cell recorded the normal force.
Two different sample geometries were used: $n=9$ rectangular cuboid samples (20 mm width, 20 mm length, 5 mm height) and $n=8$ cubic samples (side 10 mm). 
We prepared silicone samples using 3D printed moulds (0.1 mm accuracy), left to cure as per instructions and tested the same day once set.
The samples were glued between Perspex plates with quick curing cyanoacrylate (RS-Pro Radionics Ltd, Ireland).
Once the glue had cured, the Zwick Roell  tester was set to move at a rate of 0.2 mm/s for quasi-static deformation. 
Tests ended once the sample had undergone significant deformation (at least up to $K=0.35$) or the glue failed.
Finally we analysed the data with Rstudio (Integrated Development for R. RStudio, PBC, Boston, MA) to arrive at plots  of the mean data and standard deviation.

We performed a series of Finite Element simulations using ABAQUS/Standard (2017) \cite{Abaqus}. 
We used samples with 20 mm width, 20 mm length, and different heights $H$ (from 1 mm to 20 mm), with  top and bottom surfaces clamped to two elastic plates with 1 mm height, 30 mm width and 30 mm length.
The samples were assumed to be perfectly incompressible hyperelastic Mooney-Rivlin solids with the same initial shear modulus $\mu=62.5$ kPa and different $\beta$ (varying from -0.3 to 1). The plates were assumed to be linearly elastic with a Poisson ratio of 0.3 and a Young modulus of 200 GPa, so that they were effectively rigid compared to the samples.
The bottom plate was totally constrained during the deformations and the simple shear displacements were applied to the top plate and transferred to the samples. We extracted the normal force by measuring the normal reaction force of the top plate.

We used 8-node linear reduced integration brick elements (C3D8R) to discretize the elastic plates and 8-node linear reduced integration hybrid brick elements (C3D8RH) to discretize the incompressible hyperelastic samples.
To reduce the computational time and the impact of corner distortions on the simulation, we used a linearly distributed mesh, and made the corners have the densest mesh.
\md{The minimum and maximum mesh sizes in the length and width directions are both twice the minimum mesh size in the height direction. 
To prevent excessive mesh density for larger sample sizes, we also limited the maximum mesh size to be no less than 0.2. 
The total number of elements for the samples ranged from 30,000$-$120,000 depending on the geometry; for the plates it was 900.
 When the minimum grid size was equal or less than 0.02$H$, the finite element model gave a convergent normal force result.}

%
%
\section{Conclusion} 

The Poynting effect in simple shear is a well studied topic of nonlinear elasticity. 
It is known that it can be reversed in solids with fibres \cite{janmey2007negative, destrade2015dominant, horgan2021effect}, extreme strain-stiffening \cite{kanner2008extension},  slight compressibility \cite{destrade2012slight}, porosity \cite{de2016porosity}, mixture \cite{fall2022tuneable}, network vibrations \cite{baumgarten2018normal}, etc.
Here we saw that this reversal can also be achieved for all isotropic, moderately nonlinear, incompressible, homogeneous solids, simply by tuning the aspect ratio of the sample according to the curve in  Figure \ref{figure3}.

In practice, it is not easy to quantify the Poynting effect in the lab accurately, because the normal forces generated are of the second order in the amount of shear and subject to cell-load sensitivity. Moreover, the effect of block geometry is often overlooked when preparing samples, which can have unexpected consequences, such as producing unphysical model response with a cubic geometry \cite{budday2020fifty}. 

In summary, when the cancellation of normal force is desirable, as in seismic base isolation units, the aspect ratio of the block can be tweaked to minimise its effects.
Conversely, when implementing the simple shear testing protocol,  standards should be adhered to, so that artificial effects making the response stray away from the exact homogeneous solution are minimised.

\section*{Acknowledgments}  
%
We are most grateful to Dave Connelly for his technical support. This work received financial support from the Irish Research Council (James Blackwell).

\bibliography{reverse-Poynting-biblio}

\end{document}